\documentclass[aps,pre,showpacs,amsmath,amssymb,amsfonts,superscriptaddress,lengthcheck,twocolumn,longbibliography]{revtex4-1}

\usepackage{color}
\usepackage{graphicx}
\usepackage{amsthm}
\usepackage{subfigure}
\usepackage{tabu}
\usepackage{verbatim}
\usepackage{dcolumn}
\usepackage{bm}
\usepackage{epsf}
\usepackage{color}
\usepackage[colorlinks=true,citecolor=blue,linkcolor=blue,urlcolor=blue]{hyperref}
\usepackage{hhline}
\usepackage{mathtools}
\usepackage{bbold}

\newcommand{\be}{\begin{equation}}
\newcommand{\ee}{\end{equation}}
\newcommand{\ba}{\begin{align}}
\newcommand{\ea}{\end{align}}
\newcommand{\bi}{\begin{itemize}}
\newcommand{\ei}{\end{itemize}}

\newcommand{\bla}{bla\\bla\\bla\\bla\\bla}

\newcommand{\mc}[1]{\mathcal{#1}}

\begin{document}

\title{Thermodynamic optimization equalities in weak processes}

\author{Pierre Naz\'e}
\email{pierre.naze@unesp.br}

\affiliation{\it Universidade Estadual Paulista, 14800-090, Araraquara, S\~ao Paulo, Brazil}

\date{\today}

\begin{abstract}

Equalities are always a better option to be verified in an experiment than inequalities. In this work, for classical and isothermal processes, I provide thermodynamic equalities valid when the optimizations of macroscopic thermodynamic quantities are achieved in weak drivings. I illustrate their applicability by using them as a convergence criterion in the global optimization technique of genetic programming. 

\end{abstract}

\maketitle

\section{Introduction}
\label{sec:intro}

Due to inherent errors in experiments, observing inequalities between physical quantities is always a task one should avoid. In particular, optimization problems, where one should verify for instance the least possible value of a quantity, face such a difficulty, and strategies to overcome it are very welcome.

In recent times, many efforts have been made in strategies of optimization problems in non-equilibrium thermodynamics, in particular, to find the optimal protocols that minimize the thermodynamic work~\cite{deffner2020,blaber2023}. For classical and weak processes, an Euler-Lagrange equation has been proposed as a way to find such an optimal protocol~\cite{naze2022optimal}. Even though such an equation has not been explored in its physical meaning, and important information may be lacking.

In this work, based on the First Law of Thermodynamics, I present the physical meaning behind the Euler-Lagrange equation for the optimization of work, heat, and energy. Basically, when optimization is achieved, the rates of work, heat, and energy become particularly proportional to each other, for all instants of the process. Such relations are called in this work thermodynamic optimization equalities. They offer therefore a resolution to the problem of optimization since no more comparisons to verify inequalities will be made with them. 

As an application, I use them as a convergence criterion for the global optimization technique of genetic programming~\cite{duriez2017}, by comparing the difference of the values of an equality in a prescribed error. This seems more appropriate than using the naive criterion of waiting until the system does not evolve anymore~\cite{naze2023global}. In particular, the application is illustrated in the optimization problems of work, heat, and energy of the overdamped Brownian motion subjected to a moving laser or stiffening traps and white noise.

\section{Linear response theory}
\label{sec:lrt}

Consider a classical system, with a Hamiltonian $\mc{H}_S(z(t),\lambda(t))$, where $z(t)$ is a point in the phase space of the coordinates of the system and $\lambda(t)$ a control parameter. The system is weak coupled to a thermal reservoir of Hamiltonian $\mc{H}_{\mc{B}}(z(t),Z(t))$, where $Z(t)$ is a point in the phase space of the coordinates exclusively associated to the thermal reservoir. Starting in a thermal state of temperature $T$, the system performs a driving process $\lambda(t)=\lambda_0+g(t)\delta\lambda$, during a switching time $\tau$, where $g(t)$ is a protocol that obeys $g(0)=0$ and $g(\tau)=1$. The statistical distribution $\rho(t)$ of the total system evolves accordingly to the Liouville equation $\partial_t\rho=\mathcal{L}\rho$, where $\mc{L}$ is the Liouville operator. Suppose in particular that the sum of the Hamiltonian of the system and bath is given by $\mc{H} = \mc{H}_0+\lambda(t)\mc{H}_1$, where $\mc{H}=\mc{H}_{\mc{S}}+\mc{H}_{\mc{B}}$. In this case, the decomposition of the probability distribution $\rho(t)$ in terms of the first-order of $\delta\lambda$ is
\be
\rho(t)=\rho_0+\rho_1(t)\delta\lambda,
\ee
where $\rho_0$ is the associated canonical ensemble and
\be
\rho_1(t) = \int_0^{t} e^{(t-s)\mc{L}_0}\mc{L}_1\rho_0 g(s)ds,
\ee
with $\mc{L}_0(\cdot)=-\{\cdot,\mc{H}_0\}$ and $\mc{L}_1(\cdot)=-\{\cdot,\mc{H}_1\}$. In particular, we call $\phi(t)=\langle\{\mc{H}_1,\mc{H}_1(t)\}\rangle_0$ response function~\cite{kubo2012}, where $\{\cdot,\cdot\}$ is the Poisson bracket and $\langle\cdot\rangle_0$ the average along the initial equilibrium state. Also, considering the relation $\phi(t)=-\dot{\Psi}(t)$, we call $\Psi(t)$ relaxation function~\cite{kubo2012}. These quantities are going to be useful to express the macroscopic thermodynamic quantities as internal energy, work and heat in a simpler way.

\section{First Law of Thermodynamics}
\label{sec:1stlaw}

Consider the system plus the thermal reservoir. As the whole system is thermally isolated, the First Law of Thermodynamics is
\be
U_{\mathcal{S+B}}(\tau)=W(\tau),
\ee
where $U_{\mathcal{S+B}}(\tau)$ and $W(\tau)$ are respectively the variation of the total energy of the whole system and the work performed along the process. These quantities are given by
\be
U_{\mathcal{S+B}}(\tau)=\left[\int\mc{H}(\Gamma,t)\rho(\Gamma,t)d\Gamma\right]\bigg|_0^\tau
\ee
\be
W(\tau)=\int_0^\tau \int\partial_\lambda\mc{H}_{\mc{S}}(\Gamma,t)\dot{\lambda}(t)\rho(\Gamma,t) d\Gamma dt,
\ee
where $d\Gamma = dzdZ$. Rewriting it in terms of the variation of the internal energy $U_{\mc{S}}$ of the system along the process, one has
\be
U_{\mc{S}}(\tau)=W(\tau)-U_{\mc{B}}(\tau),
\label{eq:1stlaw1}
\ee
where
\be
U_{\mathcal{S}}(\tau)=\left[\int\mc{H}_{\mc{S}}(\Gamma,t)\rho(\Gamma,t)d\Gamma\right]\bigg|_0^\tau,
\ee
\be
U_{\mathcal{B}}(\tau)=\left[\int\mc{H}_{\mc{B}}(\Gamma,t)\rho(\Gamma,t)d\Gamma\right]\bigg|_0^\tau,
\ee
where the second term on the right-hand side of the equation can be identified as the heat spent along the process. Another way to identify this last quantity is by expressing the variation of the internal energy of the system along the process as
\be
U_{\mc{S}}(\tau) = \int \int_0^\tau\frac{d}{dt}\left[\mc{H}_{\mc{S}}(\Gamma,t)\right]\rho(\Gamma,t)dt d\Gamma,
\ee
where we have used the Liouville theorem. Using now $d/dt = -\mc{L}+\partial_t$, one has
\be
U_{\mc{S}}(\tau)=W(\tau)+Q(\tau),
\ee
with
\be
Q(\tau)=\int_0^\tau \int[\mc{L}\mc{H}_{\mc{S}}(\Gamma,t)]\rho(\Gamma,t) d\Gamma dt.
\ee
By comparison with Eq.~\eqref{eq:1stlaw1}, $Q(\tau)$ is the heat spent along the process.

\section{Thermodynamic optimization equalities} 
\label{sec:teowp}

Applying linear-response theory to express up to the second-order perturbation in the parameter $\delta\lambda$ the variation of the internal energy of the system, the work performed, and the heat spent, one has
\be
U_{\mc{S}}(\tau) = \delta\lambda\langle\mc{H}_1\rangle_0+\delta\lambda^2\int_0^\tau \dot{\Psi}(\tau-t)g(t)dt,
\ee
\be
\begin{split}
W(\tau) = \delta\lambda\langle\mc{H}_1\rangle_0+\delta\lambda^2\int_0^\tau \dot{\Psi}(\tau-t)g(t)dt\\
-\frac{\delta\lambda^2}{2}\int_0^\tau\int_0^\tau\ddot{\Psi}(t-t')g(t)g(t')dtdt',
\end{split}
\ee
\be
Q(\tau) = \frac{\delta\lambda^2}{2}\int_0^\tau\int_0^\tau\ddot{\Psi}(t-t')g(t)g(t')dtdt'.
\ee

\subsection{Optimization of work}

As it was presented in Ref.~\cite{naze2022optimal}, the total optimal protocol is a combination of a continuous protocol with jumps at the beginning and end of the process. In this manner, applying calculus of variations in the work in order to minimize it, one has the following Euler-Lagrange equation
\be
\int_0^\tau \ddot{\Psi}(t-t')g^*_{W}(t')dt'=\dot{\Psi}(\tau-t),
\ee
where $g^*_{W}(t)$ is the optimal protocol that will minimize the work. I remark that such optimal protocol is the continuous part of the total optimal protocol. In this manner, using the Euler-Lagrange equation and the First Law of Thermodynamics, the thermodynamic relations that define optimality are
\be
\dot{U}_{\mc{S}}(t) = 2\dot{W}^*(t) = 2\dot{Q}(t),
\ee
where $\dot{U}_{\mc{S}}(t), \dot{W}^*(t)$, and $\dot{Q}(t)$ are respectively the rates of the internal energy of the system, work performed, and heat spent along the process. Indeed
\be
\dot{U}_{\mc{S}}(t)=\delta\lambda^2\dot{\Psi}(\tau-t)g^*_{W}(t),
\ee
\be
\dot{W}^*(t)=\delta\lambda^2 g^*_{W}(t) \left(\dot{\Psi}(\tau-t)
-\frac{1}{2}\int_0^\tau\ddot{\Psi}(t-t')g^*_{W}(t')dt'\right),
\ee
\be
\dot{Q}(t)=\frac{\delta\lambda^2}{2} g^*_{W}(t)
\int_0^\tau\ddot{\Psi}(t-t')g^*_{W}(t')dt'.
\ee
Observe that such thermodynamic optimization equalities must be verified for all instants in the process of a particular switching time. One may argue that, for each one of those quantities, an integration in the variable $t$ in the duration of the process will furnish just a single quantity to evaluate in the laboratory, making the procedure easier. However, this statement holds only in one sense. To make valid in the reverse one, one may verify these integration identities for all switching times below the original one. In the end, the effort will be the same. The price to verify an equality for optimization is to verify it in a whole interval.

\subsection{Optimization of heat}

For its turn, to calculate the protocol that will minimize the heat, one must have to calculate the total optimal protocol obeying the boundary conditions. I assume that such optimal protocol is the combination of a continuous protocol with jumps, as it was in the case of the optimization of the work. Applying the calculus of variation, one has
\be
\int_0^\tau \ddot{\Psi}(t-t')g^*_{Q}(t')dt'=0,
\label{eq:eleqq}
\ee
or
\be
\dot{Q}^*(t)=\frac{\delta\lambda^2}{2}g^*_{Q}(t)\int_0^\tau \ddot{\Psi}(t-t')g^*_{Q}(t')dt'=0,
\ee
where $g^*_{Q}(t)$ is the continuous part of the optimal protocol that will minimize the heat. This result implies that the optimal heat must be null
\be
Q^*(\tau)=0,
\ee
which means that the process is adiabatic. A trivial total optimal protocol is $g^*_Q(t)=0$ with a jump at the end of the process. Is it possible that Eq.~\eqref{eq:eleqq} has a non-trivial solution for any $\tau$? The answer is negative, since $\ddot{\Psi}(t)$ is a negative kernel, and the double integral of $Q(\tau)$ is only zero if and only if $g^*_Q(t)=0$. Therefore, the only situation where the heat is minimized -- more than that, null --, occurs in sudden processes. This is reasonable since the system has no time to interact with the environment.

\subsection{Optimization of internal energy}

Finally, I apply the same reasoning used in the previous sections in the optimization of the variation of the internal energy of the system. The total protocol is considered to be composed of a continuous protocol and jumps. By applying the calculus of variations to the variation of the internal energy, one has
\be
\dot{\Psi}(\tau-t)=0.
\ee
This problem has a solution in two situations: $\Psi(t)=0$ or $\tau\rightarrow 0^+$ (it is expected that $\dot{\Psi}(t\rightarrow 0^+)=0$). The first case implies that the Hamiltonian of the system does not depend on the variation of the parameter, and, therefore, the energy is constant. This is out of our scope. For its turn, the second case one has an effective driving, where the optimal protocol will become a jump in the sudden process performed. In this case, one has
\be
\dot{U}_{\mc{S}}^*(t)=\delta\lambda^2\dot{\Psi}(\tau-t)g^*_{U}(t)=0,
\ee
where $g^*_{U}(t)$ is the continuous part of the optimal protocol, which can be any continuous function that connects $g(0)=0$ to $g(\tau)=1$. This result implies that the optimal variation of the energy of the system must be equal to a constant
\be
U_{\mc{S}}^*(\tau\rightarrow 0^+)=\delta\lambda\langle\mathcal{H}_1\rangle_0.
\ee

Let us see what are the effects of applying the optimal protocols of one quantity in the other ones. 

\subsection{Effects of optimal protocols in other quantities}

First, the optimal protocol of the work $g^*_W(t)$, when applied in the expression of heat or internal energy, will not lead to their optimal values, since the protocol is not null at all times. Indeed
\be
U_{\mc{S}}(\tau) = \delta\lambda\langle\mc{H}_1\rangle_0+\delta\lambda^2\int_0^\tau \dot{\Psi}(\tau-t)g^*_W(t)dt,
\ee
\be
W^*(\tau) = \delta\lambda\langle\mc{H}_1\rangle_0+\frac{\delta\lambda^2}{2}\int_0^\tau \dot{\Psi}(\tau-t)g^*_W(t)dt
\ee
\be
Q(\tau) = \frac{\delta\lambda^2}{2}\int_0^\tau \dot{\Psi}(\tau-t)g^*_W(t)dt.
\ee
On the other hand, using the optimal protocol that will minimize the heat, $g^*_Q(t)=0$, for any switching time, the work will not go to its optimal value, except for $\tau\rightarrow 0^+$. Indeed
\be
U_{\mc{S}}(\tau) = \delta\lambda\langle\mc{H}_1\rangle_0,
\ee
\be
W(\tau) = \delta\lambda\langle\mc{H}_1\rangle_0,
\ee
\be
Q^*(\tau) = 0.
\ee
Observe that, in this case, the thermodynamic optimization equalities trivially hold
\be
\dot{U}_{\mc{S}}(t) = 2\dot{W}(t) = 2\dot{Q}^*(t)=0.
\ee
Finally, to optimize the variation of energy of the system, one has to use any continuous protocol $g^*_U(t)$ that connects $g(0)=0$ to $g(\tau)=1$ in a sudden process. In this case
\be
U_{\mc{S}}^*(\tau\rightarrow 0^+) = \delta\lambda\langle\mc{H}_1\rangle_0,
\ee
\be
W^*(\tau\rightarrow 0^+) = \delta\lambda\langle\mc{H}_1\rangle_0,
\ee
\be
Q^*(\tau\rightarrow 0^+) = 0,
\ee
where all macroscopic quantities achieve their optimality. Observe that the thermodynamic quantities trivially still hold, as it was in the optimization of heat.

\section{Convergence criterion}

An application for thermodynamic optimization equalities is using them as a convergence criterion for the global optimization technique of genetic programming. This approach searches via evolutionary procedures the best protocol to minimize a cost functional~\cite{duriez2017}. The convergence criteria to stop the algorithm mostly naively used would be waiting for the non-evolution in the result for a significant period~\cite{naze2023global}. However, this is not a guarantee that, indeed, the search process has ended. Instead, the criterion proposed in this work stops the code since some thermodynamic equality is achieved with a previously established error.

The basic idea is to consider the error function
\be
\epsilon_N=\frac{1}{N}\sum_{n=0}^{N-1}\frac{|\dot{U}_{\mc{S}}(n\tau/N)-2\dot{Q}(n\tau/N)|}{\delta\lambda^2\Psi_0/\tau_R},
\ee
which will measure the average discrepancy between the quantities in the thermodynamic equality $\dot{U}_{\mc{S}}=2\dot{Q}$, and compare it with a previously established value $\epsilon_0$, at each interaction of the algorithm. If $\epsilon_N<\epsilon_0$, the best protocol at that point is chosen and the search algorithm is stopped. The error function can be redefined for other differences of the quantities in some thermodynamic equality and can be used in any optimization process of work, heat, or variation of energy of the system.

\subsection{Example: Overdamped Brownian motion}

Consider an overdamped Brownian motion subjected to a moving laser or stiffening traps and white noise~\cite{naze2022optimal}. The relaxation function associated with this system is
\be
\Psi(t)=\Psi_0 e^{-\frac{|t|}{\tau_R}},
\ee
where $\Psi_0$ is a constant and $\tau_R$ is its relaxation time. The optimal protocol of the work is~\cite{naze2022optimal}
\be
g^*_W(t)=\frac{t+\tau_R}{\tau+2\tau_R},
\ee
with jumps at the beginning and final point of the process. Let us apply the global optimization technique of genetic programming with the convergence criterion proposed to optimize the work, heat and variation of energy of the system.

\begin{figure}[t]
    \centering
    \includegraphics[scale=0.27]{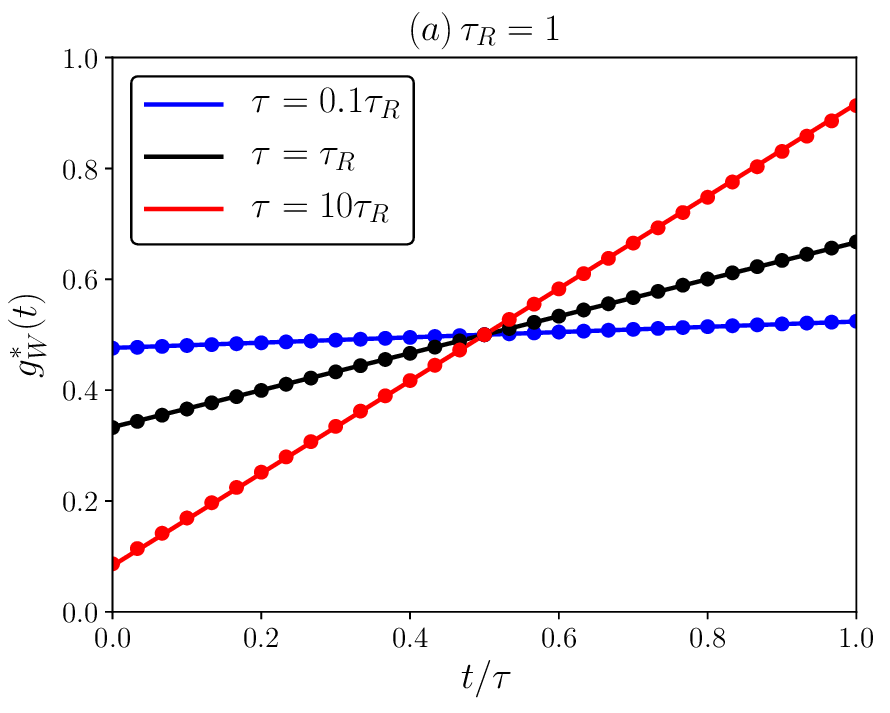}
        \includegraphics[scale=0.27]{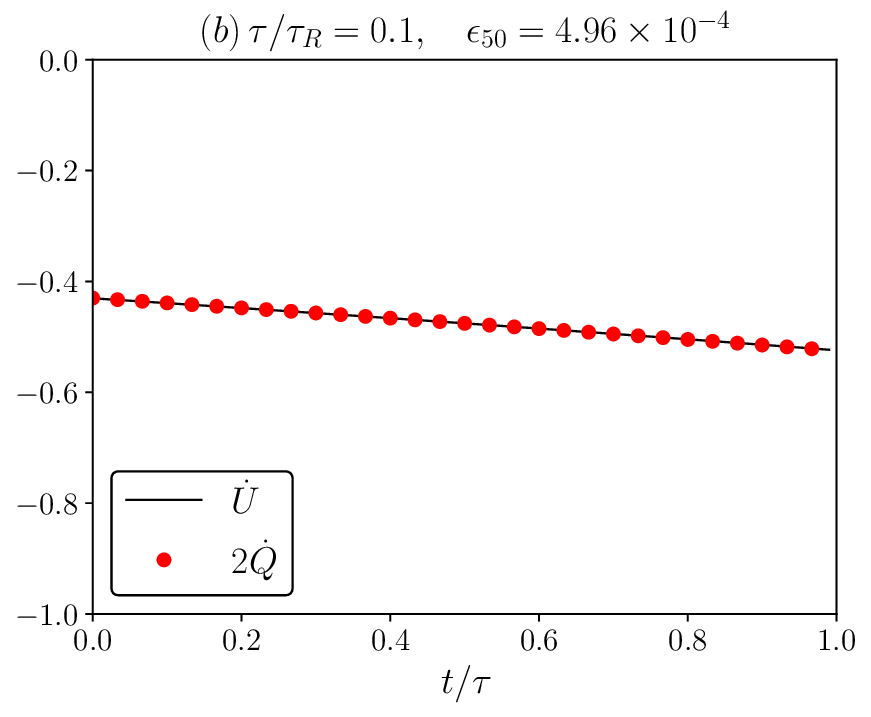}\\
            \includegraphics[scale=0.27]{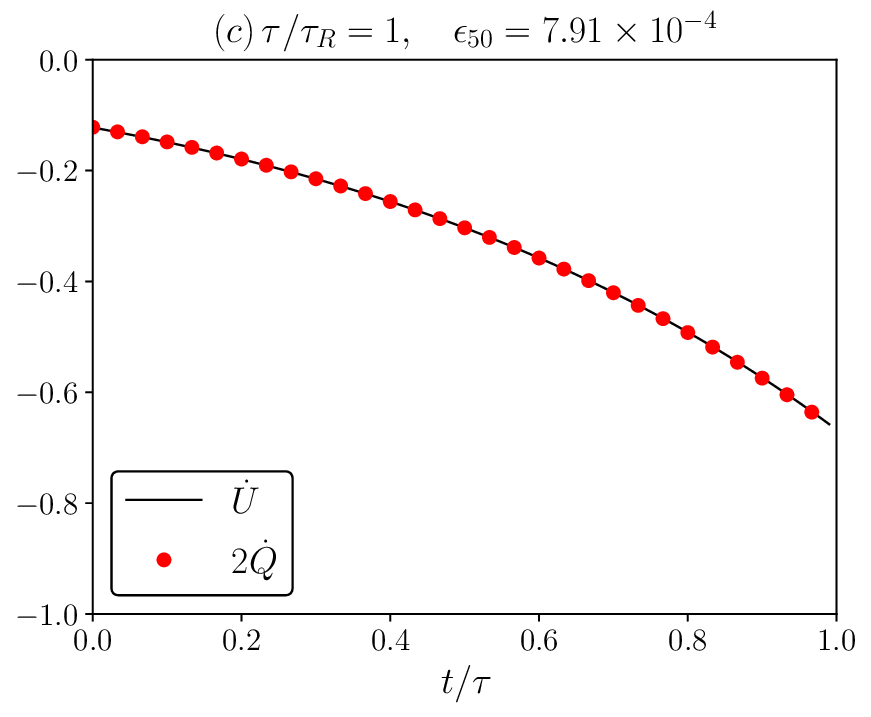}
                \includegraphics[scale=0.27]{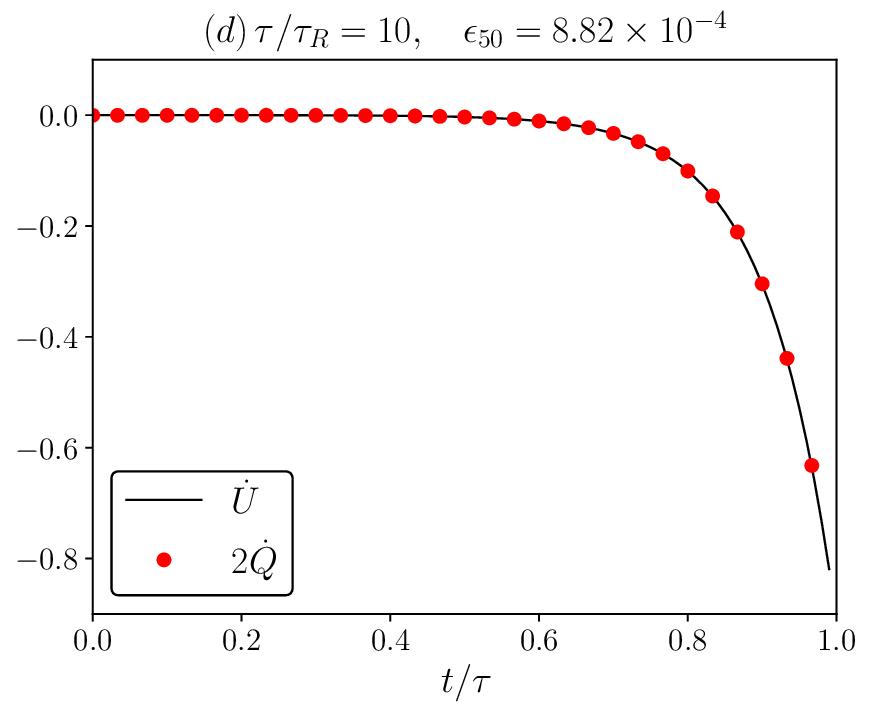}
    \caption{Graphic (a): Comparison between numerical optimal protocols for work and exact result for switching times $\tau/\tau_R=0.1,1,10$. Graphics (b)-(d): Comparison between $\dot{U}_{\mc{S}}$ and $2\dot{Q}$ for different switching times at the moment where the code stopped according to our convergence criterion. They are measured in unit $\Psi_0 \delta\lambda^2/\tau_R$.}
    \label{fig:fig1}
\end{figure}

For the optimization of work, considering $N=50$ and $\epsilon_0=10^{-3}$, graphic (a) of Fig.~\ref{fig:fig1} depicts the comparison of the outcomes of the code with the exact result for switching times $\tau=0.1\tau_R, \tau_R, 10\tau_R$. The agreement is satisfactory, and, in the case of the work, the running time of the code reduces sensibly in comparison with the numerical experiments of my previous work~\cite{naze2023global}. Graphics (b)-(d) of Fig.~\ref{fig:fig1} show the comparison between $\dot{U}_{\mc{S}}$ and $2\dot{Q}$ at the moment when the code stopped for different switching times. The value of the error $\epsilon_{50}$ is presented for each graphic as well.

\begin{figure}[t]
    \centering
    \includegraphics[scale=0.27]{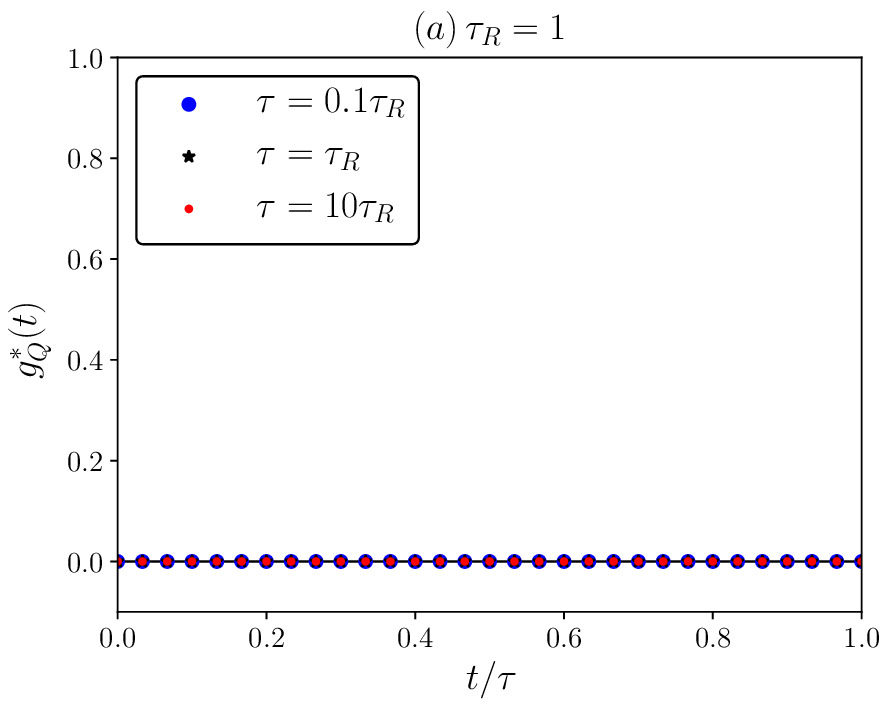}
        \includegraphics[scale=0.27]{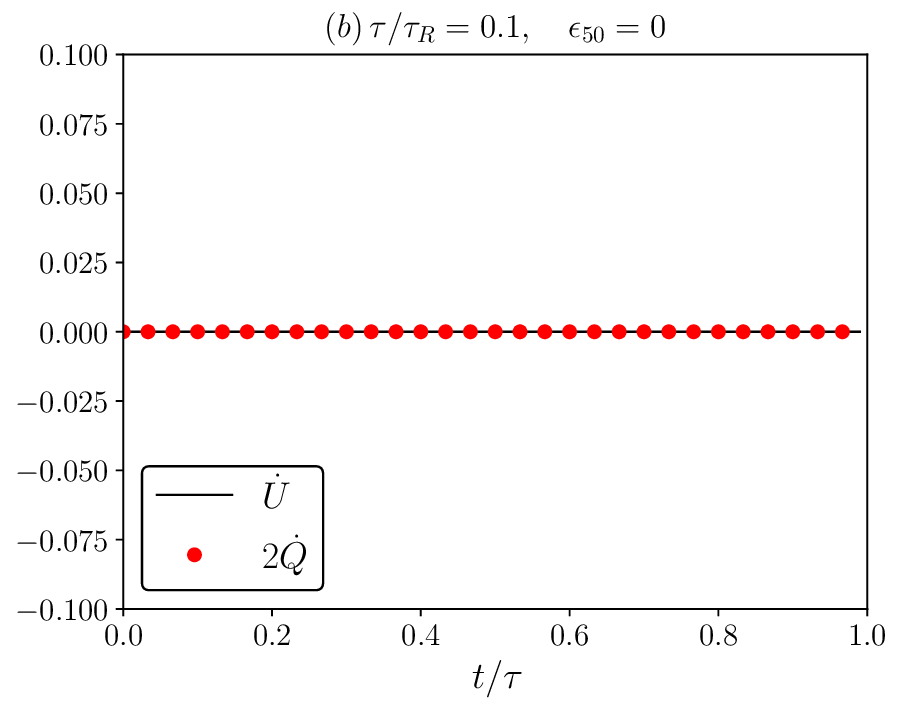}\\
            \includegraphics[scale=0.27]{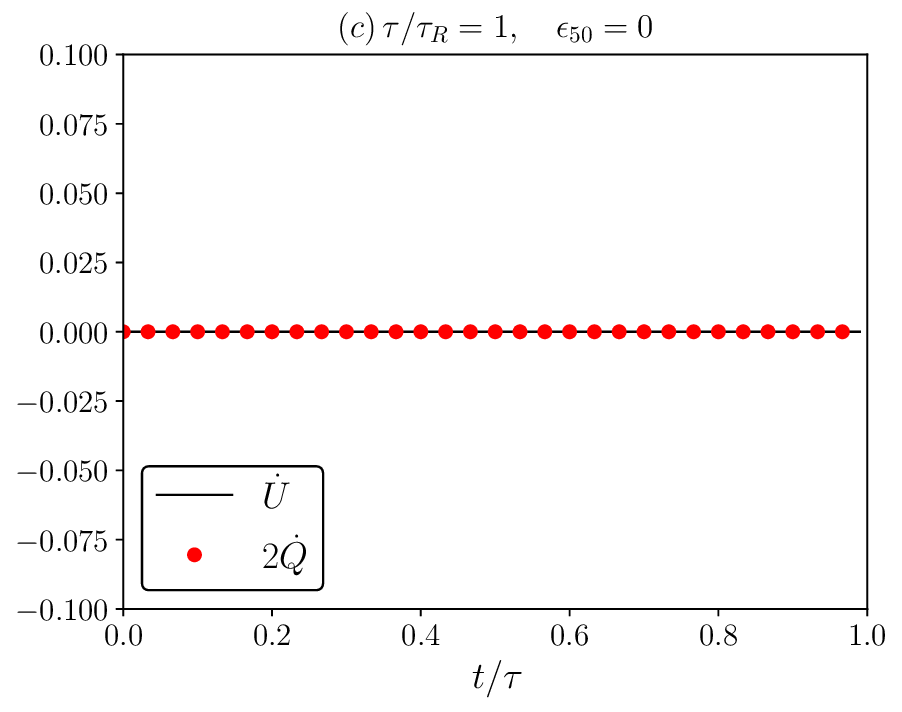}
                \includegraphics[scale=0.27]{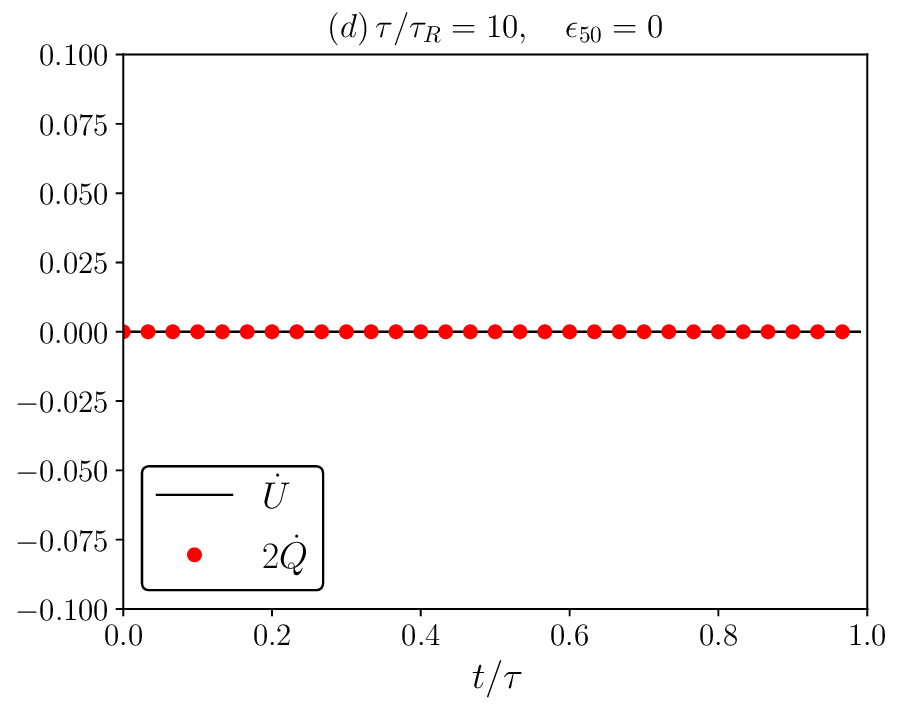}
    \caption{Graphic (a): Comparison between numerical optimal protocols for heat and exact result for switching times $\tau/\tau_R=0.1,1,10$. Graphics (b)-(d): Comparison between $\dot{U}_{\mc{S}}$ and $2\dot{Q}$ for different switching times at the moment where the code stopped according to our convergence criterion. They are measured in unit $\Psi_0 \delta\lambda^2/\tau_R$.}
    \label{fig:fig2}
\end{figure}

For the optimization of heat, considering $N=50$ and $\epsilon_0=10^{-3}$, graphic (a) of Fig.~\ref{fig:fig2} depicts the comparison of the outcomes of the code with the exact result for switching times $\tau=0.1\tau_R, \tau_R, 10\tau_R$. The agreement is satisfactory. Graphics (b)-(d) of Fig.~\ref{fig:fig2} show the comparison between $\dot{U}_{\mc{S}}$ and $2\dot{Q}$ at the moment when the code stopped for different switching times. The value of the error $\epsilon_{50}$ is presented for each graphic as well.

On the other hand, for the optimization of the variation of energy of the system, performed always in sudden process, the outcomes show that there is no optimal protocol for this system. Indeed, the code runs indefinitely, always reducing the variation of energy of the system. This occurs because the derivative of the relaxation function is discontinuous at $t=0$, achieving the limit
\be
\dot{\Psi}(t\rightarrow 0^+)=-\frac{1}{\tau_R}\neq 0,
\ee
excluding therefore the necessary hypothesis $\dot{\Psi}(t\rightarrow 0^+)=0$.

\section{Final remarks} 

In this work, I present the physical meaning of the Euler-Lagrange equations for the optimization of work, heat, and energy, calling them thermodynamic optimization equalities. Basically, the rates of the work, heat, and energy are particularly proportional between themselves when optimization is achieved. As an application, I proposed as a convergence criterion the achievement of the thermodynamic optimization equalities with prescribed error in the global optimization technique of genetic programming. I illustrate it with the overdamped Brownian motion, observing agreement for work and heat. Surprisingly, the process of the optimization of energy shows the non-existence of an optimal protocol. For future studies, important questions are: Do the thermodynamic optimization equalities break down for higher orders? How do they behave in the slowly-varying regime and arbitrarily strong processes?

\begin{acknowledgments}

I thank Maurice de Koning for the question made during my doctorate defense that motivates the applicability of this work, and Marcus V. S. Bonan\c{c}a for enlightening discussions.

\end{acknowledgments}

\bibliography{bibliography}
\bibliographystyle{apsrev4-2}

\end{document}